\documentclass[twocolumn]{emulateapj}
\journalinfo{{\sc Accepted by The Astrophysical Journal}}
\usepackage{natbib}
\usepackage{amssymb}
\usepackage{graphicx}
\usepackage{apjfonts}

\newcommand{\chan}{\emph {Chandra}}
\newcommand{\rxte}{\emph {RXTE}}

\begin{document}
\title{{\em Chandra} and {\em RXTE} Spectra of the Burster GS 1826--238}
  
\author{Thomas W. J. Thompson\altaffilmark{1},
Richard E. Rothschild\altaffilmark{1},
John A. Tomsick\altaffilmark{1}, 
Herman L. Marshall\altaffilmark{2}}

\altaffiltext{1}{Center for Astrophysics and Space Sciences, University
of California, San Diego, La Jolla, CA 92093; email: tthompson@physics.ucsd.edu }
\altaffiltext{2}{Center for Space Research, 77 Vassar Street, Massachusetts 
Institute of Technology, Cambridge, MA 02139}

\begin{abstract}
Using simultaneous observations from the {\em Chandra X-Ray Observatory}
and the {\em Rossi X-Ray Timing Explorer}, we investigated the low-mass 
X-ray binary (LMXB) and ``clocked burster'' GS 1826--238 with the goal 
of studying its spectral and timing properties. The uninterrupted 
\chan~observation captured 6 bursts (\rxte~saw 3 of the 6), yielding a 
recurrence time of 3.54 $\pm$ 0.03 hr. Using the proportional counter 
array on board {\em RXTE}, we made a probable detection of 611 Hz burst 
oscillations in the decaying phases of the bursts with an average rms 
signal amplitude of 4.8\%. The integrated persistent emission spectrum 
can be described as the dual Comptonization of $\sim$ 0.3 keV soft 
photons by a plasma with $kT_{\rm e} \sim 20$ keV and $\tau \sim 2.6$ 
(interpreted as emission from the accretion disk corona), plus the 
Comptonization of hotter $\sim$ 0.8 keV seed photons by a $\sim$ 6.8 keV 
plasma (interpreted as emission from or near the boundary layer). We 
discovered evidence for a neutral Fe K$\alpha$ emission line, and we 
found interstellar Fe L {\sc II} and Fe L {\sc III} absorption features. 
The burst spectrum can be fit by fixing the disk Comptonization 
parameters to the persistent emission best-fit values, and adding a 
blackbody. The temperature of the boundary layer seed photons was tied 
to the blackbody temperature. The blackbody/seed photon temperature at the 
peak of the burst is $\sim$ 1.8 keV and returns to $\sim$ 0.8 keV over 200 
s. The blackbody radius is consistent with $R_{\rm bb} \approx$ 10.3--11.7 
km assuming a distance of 6 kpc, though this value cannot be interpreted 
as the physical size of the neutron star due to partial covering of the
stellar surface by the accretion disk. By accounting for the fraction of 
the surface that is obscured by the disk as a function of binary 
inclination, we determined the source distance must actually be near 5 kpc 
in order for the stellar radius to lie within the commonly assumed range 
of 10--12 km. The order of magnitude increase in flux at burst peak is 
seen to cause Compton cooling of the electron plasma surrounding the disk, 
as the plasma temperature decreases to $\sim$ 3 keV at burst onset, and 
then slowly returns to the persistent emission value after about 150 s.
\end{abstract}

\section{Introduction}
The low-mass X-ray binary (LMXB) GS 1826--238
was discovered with \emph{Ginga} in 1988 (Makino et al. 1988). Due to its temporal
and spectral similarities to Cyg X--1 and GX 339--4, the source was originally 
tentatively classified as a black hole candidate (BHC) (Tanaka 1989). Later optical 
studies lead to the identification of a $V=19.3$ mag (and therefore low-mass) optical
counterpart (Barret et al. 1995). The companion was subsequently found to have a 2.1 
hr modulation (and implied orbital period) and a refined position of $\alpha=18^h
29^m28\fs2$ and $\delta=-23\arcdeg47\arcmin49\farcs12$ (J2000) (Homer et 
al. 1998). 

The spectral and temporal characteristics that Tanaka (1989) initially used to associate 
GS 1826--238 with a black hole system were later found to be present in other X-ray 
bursters (e.g., 4U 1608--522, Yoshida et al. 1993). Moreover, the photon index of its 
energy spectrum was measured to have a relatively low cut-off energy ($\sim$ 58 keV) for
a BHC, and was perhaps more indicative of the typically cooler neutron star (NS) hard
X-ray spectra (Strickman et al. 1996). X-ray bursts from this source were first 
conclusively observed with \emph{BeppoSAX} by Ubertini et al. (1997), firmly establishing 
the source as a NS and strongly suggesting it to be weakly magnetized ($B < 10^{10}$ G). 

The periodicity of the type I bursts from GS 1826--238 has been remarkably stable
over the span of years (Ubertini et al. 1999). Although quasi-periodic bursting is not 
unique among LMXBs, such consistency over long durations is indeed unusual. The regular
intervals between bursts suggest that the accretion rate is stable, that the accreted matter is 
completely consumed during the bursts, and that the fraction of the stellar surface covered 
prior to each burst is approximately constant. Investigations of burst recurrence rates 
and energetics have lead to convincing arguments that type I bursts stem from unstable 
thermonuclear burning of accreted hydrogen and helium (e.g., Strohmayer \& Bildsten 2003). 
As freshly accreted material falls onto the NS surface, it is hydrostatically compressed 
by new material at a rate per unit area $\dot{m} \sim$ 10$^{4}$ g cm$^{-2}$ s$^{-1}$, 
assuming isotropic accretion and a NS radius of 10 km. The thermal energy deposited by the 
infalling matter causes temperatures in most of the thin NS ``atmosphere'' to exceed $10^7$ 
K, so that during the accumulation phase hydrogen burns via the hot CNO cycle at a rate that is 
limited only by 
the mass fraction $Z_{\rm CNO}$ and not the temperature (Bildsten 2000). Within hours to days, 
the extreme gravity on the NS surface ($\sim$ 10$^{14}$ cm s$^{-2}$) compresses the 
accumulated matter to densities high enough to trigger unstable thermonuclear ignition. 
GS 1826--238, in particular, has near limit-cycle behavior with stable hydrogen burning during 
the accumulation phase followed by mixed hydrogen and helium burning triggered by 
thermally unstable helium ignition (Bildsten 2000). The $\alpha$-parameter -- the ratio of the 
integrated persistent fluence between bursts to the burst fluence -- and the long burst
duration ($\sim$ 150 s), imply that after thermonuclear ignition the hydrogen burns via 
the rapid-proton ($rp$) process where energy is released through successive proton captures 
and $\beta$ decays (Wallace \& Woosley 1981). The measured $\alpha$-value for GS 1826--238 
of $\sim$ 42 (Galloway et al. 2004; hereafter G04) is remarkably consistent with 
theoretical predictions: The gravitational energy released during accretion onto a 
1.4 M$_{\odot}$ NS is about 200 MeV per nucleon, while the energy released through 
thermonuclear fusion is about 5 MeV per nucleon for a solar mix going to iron group elements, 
giving an expected value of 40. Moreover, the variation in $\alpha$ with the global 
accretion rate $\dot{M}$ implies solar metallicity in the accreted layer (G04),
although recent work using an adaptive nuclear reaction network shows that the critical mass 
required for a burst is independent of the composition of the accreted material (Woosley et al. 
2004), and so attempts to infer the metallicity of the fuel from burst properties is complicated.

During accretion the gravitational potential energy released per nucleon is deposited partly 
in the disk, and partly in the boundary layer or on the NS surface. In low accretion rate (atoll)
LMXBs, the dominant persistent emission spectral component is clearly 
inverse Comptonization, where soft ``seed'' photons emitted from a cold layer in the 
accretion disk gain energy in successive scatterings with hot 
electrons in the disk or surrounding accretion disk corona (ADC) (Church 2001). Stable 
hydrogen burning between bursts and the thermal energy released on impact are observable
as blackbody emission, assuming there is an optical path through which the radiation can 
escape. The addition of a blackbody component in spectral models is not always necessary (as 
we find in \S~\ref{spec}), which suggests this form of radiation is (up)scattered or absorbed 
by the obscuring disk or ADC. Another possibility is that the blackbody component is simply 
too weak to be observed relative to the Comptonized emission. 

Mitsuda et al. (1989) proposed that most LMXB spectra can be described by a 
multi-temperature disk blackbody from the inner accretion disk plus blackbody emission 
from the NS which is Comptonized in the local region of the star. More recently, 
the dipping class of LMXBs has constrained the types of emission models that are 
most likely. The constraints are provided by the fact that the models must fit the 
persistent emission at various levels of decreased flux resulting from 
partial occultation and absorption of the NS, accretion disk, and corona by the bulge 
in the outer disk where accretion flow from the companion star impacts. 
The evolution of these spectra at various levels of flux are 
well-described by a model consisting of point-like blackbody emission plus 
Comptonized emission from an extended accretion disk corona (e.g., Church 
et al. 1997, 1998; Ba\l uci\'{n}ska-Church et al. 2001; Smale et al. 
2001). During dipping, the blackbody component is observed to disappear rapidly, 
indicating that the emission comes from a point and not from an extended region. 
As expected, the Comptonized component is observed to gradually decrease.
By measuring the ingress times of the dip, Church (2001) was able to 
estimate the size of the Comptonizing region for several sources and found the 
radius of the region to be typically $\sim$ 50,000 km. This measurement is 
consistent with the value obtained much earlier by Canizares (1976) for 3U 1820--30.
The large extent of the emission region is evidence for a Comptonized component that 
is emitted from an ADC and not the disk itself. Moreover, Smale et al. (2001) 
observed complete covering of this region in many sources, suggesting that the 
corona must be geometrically thin, as it would be unlikely that a corona with 
spherical geometry would be completely occulted. On the contrary, studies of 
emission and absorption features in dipping sources by Boirin et al. (2005) show 
that the spectra are also consistent with a less strongly ionized absorber along 
the line of sight rather than a simple increase in absorption. Progressive covering 
of the Comptonized component in their models is not required.

The persistent emission spectrum of GS 1826--238 has been discussed by previous 
authors (e.g., Barret et al. 1995, Ubertini et al. 1999, Del Sordo et al. 1999,
in 't Zand et al. 1999a, Barret et al. 2000, Kong et al. 2000). The spectral model that 
has consistently produced an acceptable fit is a blackbody plus Comptonized emission, 
the latter being modeled in {\sc xspec} with either a cut-off power law ({\tt
cutoffpl}) or the more explicit {\tt comptt}. Broadband \emph{BeppoSAX} spectra in the 
0.1--200 keV range found the spectrum to be consistent with the Comptonization of 
a 0.6 keV Wien spectrum by a plasma with $kT_{\rm e} \sim 20.7$ keV, plus a 
3.8 keV blackbody (in 't Zand et al. 1999a). Del Sordo et al. (1999) found the spectrum 
to be well-fitted with a blackbody plus cut-off power law, with $kT_{\rm bb} \sim 0.9$ keV, 
$\Gamma \sim 1.3$, and cut-off energy $\sim 50$ keV. Strickman et al. (1996) and in 't 
Zand et al. (1999a) also obtained a fit with this model and the results of each group 
are fairly consistent. Kong et al. (2000) fit the persistent emission spectrum 
from 0.5--10 keV with a blackbody ($\sim 0.7$ keV) plus power law ($\Gamma \sim 1.1$). 

The burst emission spectrum has been studied a few times previously. in 't Zand et al. 
(1999a) studied burst spectra and found significant flux up to 60 keV, indicating that
the burst emission may be Comptonized in a similar manner as the persistent emission. The peak 
blackbody temperature was measured to be $\sim$ 2 keV, which cooled to about 1.3 keV over 
100 s. Ubertini et al. (1999) studied the spectrum for two bursts, though with only two 
separate integration bins of 13 s at burst peak and 43 s through the decay. The peak spectrum 
was fit with a blackbody temperature of $\sim$ 2.1--2.3 keV and the decay with temperature 
$\sim$ 1.6--1.9 keV. Kong et al. (2000) fit the burst spectrum by fixing the power law 
component to its persistent value and adding a blackbody; the peak temperature was $\sim 
2.6$ keV. Marshall et al. (2003) searched the present data for absorption features during
the burst peak and decay to search for evidence for the gravitational redshift to the 
surface of the NS. No features were observed. 

We begin in \S~2 by presenting the observations of GS 1826--238 that were used in this study, 
and we specifically discuss the data preparation prior to spectral analysis. 
In \S~\ref{bp} we measure the burst periodicity and compare to previous measurements. We use
the burst recurrence time and energetics to discuss the type of burning occurring on the 
NS surface. We conclude the section by discussing a probable detection of burst 
oscillations during the decaying phases of the burst. 
In \S~\ref{spec} we examine the evolution of the broadband persistent emission spectrum
between bursts to see whether or not any significant changes in the parameters occur
during the accumulation phase. We find that the persistent emission data from
0.5--200 keV are best fit with a model characterized by absorbed emission from two distinct
Comptonizing regions, plus iron line emission. We also report the detection of interstellar 
absorption features. We study the evolution of the burst spectrum in \S~\ref{burspec} by 
adding a blackbody to the persistent emission spectrum. The disk Comptonization parameters 
are fixed at their persistent emission best-fit values, and the temperature of the boundary 
layer seed photons is tied to the blackbody temperature. From the blackbody parameters we 
derive a blackbody radius, and by accounting for the accretion disk covering factor, we 
constrain the source distance assuming a 10--12 km NS radius. In \S~\ref{geo} we attempt to 
obtain information on the geometry of the burning by comparing the persistent and burst 
spectra. We conclude by summarizing the major results or our work. 

\section{Observations and Analysis}
In this paper we utilize two simultaneous observations from 2002 July 29, taken with 
\chan~and {\em Rossi X-Ray Timing Explorer}~({\em RXTE}). The 68.2 ks \chan~observation 
was made using the Advanced CCD Imaging 
Spectrometer (ACIS; Garmire et al. 2003) and is sensitive to photons from 0.3--10 
keV. The source was focused onto one of the 
back-illuminated ACIS chips (S3). To minimize the adverse effects due to pile-up, 
and to gain better time resolution on the rise of the bursts, we used half-frame 
readout on the inner 4 ACIS chips with 1.74 s frame-time. We use the ``level 2'' 
event lists from the standard data processing, and apply a standard correction to 
destreak the ACIS-S4 chip caused by a flaw in the chip readout. To obtain high 
resolution spectra, we had the High Energy Transmission Grating (HETG; 
Canizares et al. 1992) inserted into the optical path.
 
The 36 ks \rxte~observation uses the Proportional Counter Array (PCA; Jahoda 
et al. 1996), and the High Energy X-Ray Timing Experiment (HEXTE; Rothschild et 
al. 1998). The PCA is made up of five proportional counter units (PCUs) and is
sensitive to photons from 2--60 keV. The HEXTE instrument comprises 
two clusters, each of which contains four NaI/CsI scintillation detectors, and
is sensitive to photons from 15--250 keV. Both instruments have large effective 
areas ($\sim$ 6000 cm$^{2}$ and 1400 cm$^{2}$, respectively) and microsecond timing.
Table 1 provides a summary of the GS 1826--238 observations used in the spectral analysis.
\begin{deluxetable}{clcr}
\tablenum{1}
\tablecolumns{4}
\tabletypesize{\scriptsize}
\tablewidth{0pt}
\tablecaption{\sc{{\em Chandra} and {\em RXTE} Observations of GS 1826-238}}
\tablehead{
\colhead{Obs. ID     }  & 
\colhead{Telescope/  }  & 
\colhead{Energy Band }  & 
\colhead{Exp.        }  \\
\colhead{ } &
\colhead{Instrument  } & 
\colhead{(keV)} & 
\colhead{(ks) }
}
\startdata
2739 & \chan /ACIS-S & 0.5--8.3 & 68.2 \\ 
01,02,000\tablenotemark{a} & \rxte /PCA & 3--23 & 24.6\tablenotemark{b} \\ 
01,02,000\tablenotemark{a} & \rxte/HEXTE & 17--200 & 19.4\tablenotemark{b} \\ 
\enddata
\tablecomments{All observations are from 2002 July 29.}
\tablenotetext{a}{The \rxte~observation IDs are each preceded by 70044-01-01-}
\tablenotetext{b}{Due to earth occultation and the satellite's passage
through the South Atlantic Anomaly, there are time gaps of 15--30 min in 
coverage; the times listed represents the sum of 9 separate observation
intervals.}
\end{deluxetable}
\subsection{Data Preparation for Spectral Analysis} \label{dpfsa}
The \chan~HETG Spectrometer is composed of the Medium Energy Grating (MEG)
and the High Energy Grating (HEG). The two gratings are slightly offset
and appear as ``whiskers'' traversing all ACIS chips. The MEG is calibrated
from 0.3--5.0 keV and the HEG is calibrated from 1.0--8.3 keV. In this 
analysis, we only made use of the 1$^{\rm st}$ order MEG/HEG data. All of the 
\chan~analysis made extensive use of the Chandra Interactive Analysis of 
Observations (CIAO) version 3.01 software with calibration version (CALDB) 2.26.
We used the CIAO routine \emph{tgextract} for the extraction of grating spectra, 
\emph{mkgrmf} to produce response matrices, and \emph{fullgarf} to create the 
auxiliary response matrices. We added the positive and negative diffraction 
orders and their corresponding auxiliary response files using the script 
\emph{add\_grating\_orders}. Systematic errors of 10\% were derived from 
calibration observations of the Crab pulsar and Mrk 421, and were added to both 
the persistent and burst emission fits.\footnote{see http://space.mit.edu/ASC/calib/hetgcal.html} 
We binned the persistent and burst emission data in the largest possible bins to 
obtain maximum statistical quality, while maintaining sufficient resolution to 
observe prominent absorption or emission lines. For the persistent 
emission analysis we bin the data in 1000 count PHA bins, and in the burst 
analysis we use 500 count bins. This generally led to bins with width 
$\sim$ 50--100 eV. We excluded the MEG 1$^{\rm st}$ order data 
from 0.8--0.9 keV because these data fall on a gap between the ACIS chips. 
After preliminary analysis, we concluded that the data near 2.1 keV are 
affected by a residual calibration uncertainty in the response due to an iridium 
M-edge. To correct for this, we included an inverse edge (e.g., Miller et 
al. 2002), frozen at 2.065 keV though allowing the optical depth to vary. The best-fit 
optical depth was consistently $\sim -0.16$. In addition, the data from 2.0--2.5 keV 
are possibly affected by pile-up,
which we estimate to be about 10\% at the maximum. Spectral distortion due to
pile-up is accounted for by the HETG systematics.

The extraction of PCA and HEXTE energy spectra used scripts developed at the 
University of California, San Diego and the University of T\"{u}bingen that 
incorporate the standard software for \rxte~data reduction (FTOOLS).
The large effective area of the PCA provides excellent statistics without
rebinning the data, however the first three channels and the data 
above 23 keV were not used. We placed systematic uncertainties of 0.5\% on the 
PCA data to account for the likely uncertainty in the response (Kreykenbohm et al. 
2004). The HEXTE data 
below 17 keV and above 200 keV are not included in the analysis. We used 
successively larger energy bins at higher energies to account for the decreasing 
flux density. During the burst intervals
HEXTE did not yield a significant number of counts above 40 keV. We included no
systematics to the HEXTE data. For analysis during the burst time intervals 
we added the data for the 3 bursts (corresponding to the 
2$^{\rm nd}$, 3$^{\rm rd}$, and 6$^{\rm th}$ \chan~bursts) viewed when PCUs 0 
and 2 were taking data; PCUs 0 and 2 had the greatest coverage in this observation. 
The PHA bins were added together using the FTOOLS task \emph{mathpha}.

\section{Burst Periodicity and Timing Characteristics} \label{bp}
As discussed in \S~1, GS 1826--238 has exhibited a nearly periodic bursting behavior 
since type I bursts from this source were first discovered in 1997. The remarkable 
consistency of the burst cycle can easily be seen in Fig. 1. The \chan~light 
curve uses all HETG data from 0.3--8 keV, but does not include any counts
from the zeroth order source region as these data are affected by severe pile-up.
Not only is the burst recurrence time approximately constant, but the light curve of any
particular burst is virtually indistinguishable from any other (see Fig. 1 ({\em c})).  
Although the time between sequential bursts has been rather constant, the average burst 
recurrence time has been observed to be steadily decreasing over long time scales 
(Cocchi et al. 2000, G04). Using a total of 44 \rxte~observations from 1997 November to 
2002 July, G04 measured
the burst interval to be 5.74 $\pm$ 0.13 hr initially, 4.10 $\pm$ 0.08 hr in 
2000, and 3.56 $\pm$ 0.03 hr in 2002. We measured the average burst recurrence time 
of the five complete \chan~intervals to be 3.54 $\pm$ 0.03 hr (12750 $\pm$ 102 s; also in 
2002). The individual intervals were measured from burst peak to peak and are listed in 
Table 2. The 2002 recurrence time measurement by G04 was measured using the same PCA data 
that we use in this paper, while our measurement is based on the \chan~data. 

\begin{figure*}[h] \label{fig1}
\centering
\includegraphics[width=5in]{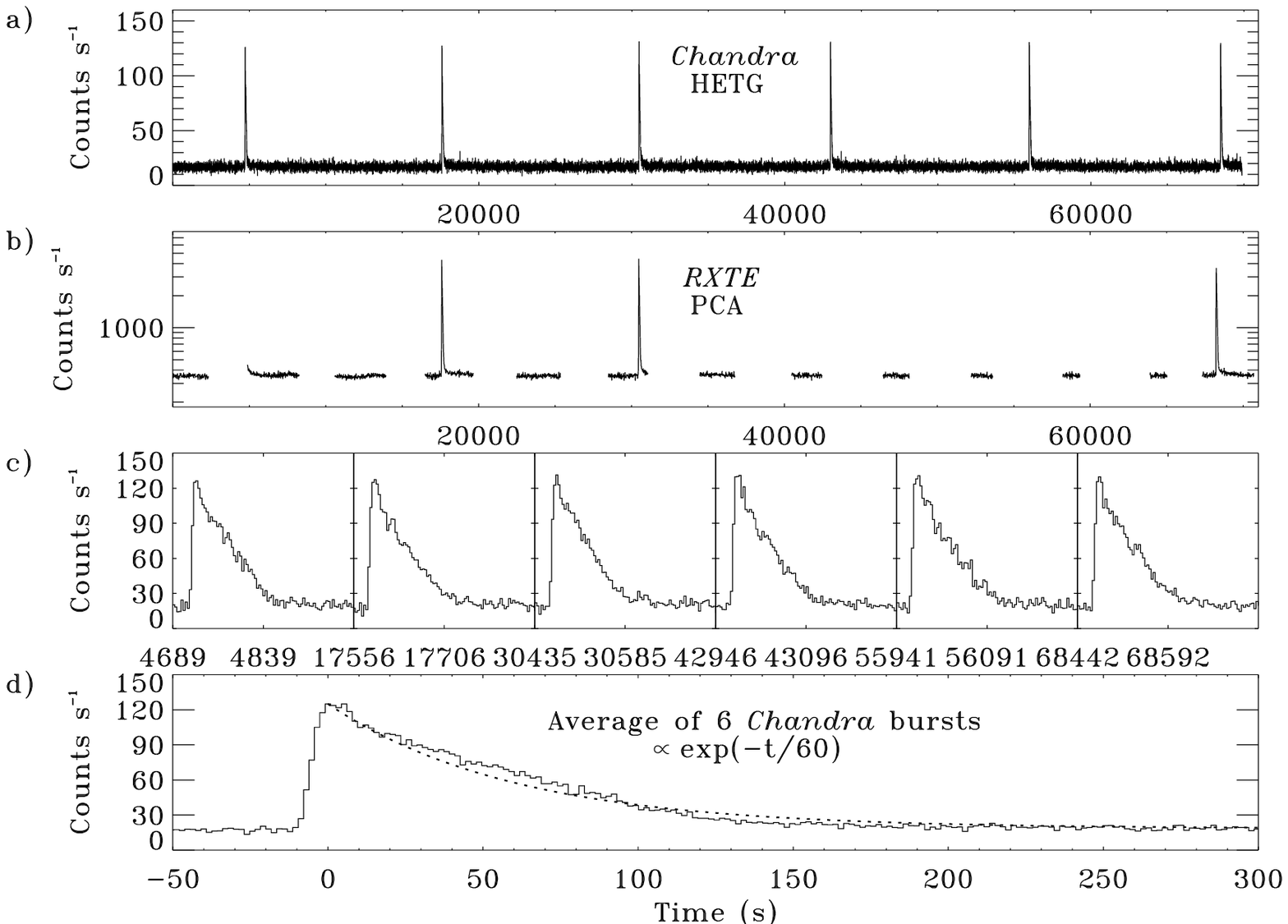}
\caption{Light curves for the {\em Chandra} HETG and {\em RXTE} PCA instruments. The top 
two light curves ({\em a} and {\em b}) show the intervals when each telescope was taking
data. The \chan~light curve includes the HETG data from 0.3--8 keV and has time resolution
of 3.48 s, as does the data in plot ({\em c}). The PCA light curve includes data 
from 3--8 keV and uses 16 s time bins. The time axis of the top three plots begins at 
MJD 52484.2039. The lower plots show the individual ({\em c}) and average ({\em d}) of 
the 6 \chan~bursts. We used the 
smallest possible binning of 1.74 s, and the curves were aligned using the rises of the 
bursts. The zero point reference time for plot ({\em d}) also corresponds to the peak of the
burst. The curve can be approximated by a decaying exponential with scale time 60 s, however 
this value cannot be interpreted physically since it includes all instrumental effects 
resulting from differences in the detector response with energy.}
\end{figure*}
\begin{deluxetable}{cc}[!b] 
\tablenum{2}
\tabletypesize{\scriptsize}
\tablecolumns{2}
\tablewidth{0pt}
\tablecaption{\sc{Time Between Bursts}} 
\tablehead{
\colhead{Burst Interval} &
\colhead{$\Delta t$ (s)}
}
\startdata
1$\longrightarrow$2 & 12861 \\
2$\longrightarrow$3 & 12884 \\
3$\longrightarrow$4 & 12510 \\
4$\longrightarrow$5 & 12994 \\
5$\longrightarrow$6 & 12502 \\ 
\enddata
\tablecomments{The individual burst intervals were measured from the peak of one
burst to the next. There is a small uncertainty in these measurements since the
\chan~data have a frame time (and time resolution) of 1.74 s.} 
\end{deluxetable}

The 40\% decrease in the burst recurrence time from 1997 to 2002 has been
coupled with a 66\% increase in the mean persistent flux (G04). Such behavior
is expected since the increase in persistent luminosity is assumed to be due
to an increase in the global accretion rate, and therefore less time is required
to reach the critical amount of fuel. While the 
observations of G04 are consistent with this trend, Bildsten (2000) found 
exactly the opposite behavior in many low accretion rate LMXBs. For example,
bursts from 4U 1705--44 became less frequent as $\dot{M}$ increased. To 
explain the conundrum, Bildsten (2000) suggested that a greater fraction of 
the stellar surface of 4U 1705--44 may be covered prior to ignition, so that 
the accretion rate per unit area actually decreased. This may also be the case 
for GS 1826--238: If the relation between the rise in the persistent flux and 
the decrease in the burst recurrence time were linear, we would expect an 
even greater decrease in burst recurrence time than has been observed by G04.
Another possible explanation for the longer-than-expected recurrence time regards
the fact that at smaller recurrence times, less helium will be made prior to 
ignition, and so the column depth required to produce the instability will increase 
(Cumming \& Bildsten 2000).

The location of the burning during bursts is not assumed to be spherically
symmetrical. It is likely that ignition begins near the stellar equator 
since matter is preferentially deposited there during accretion (Spitkovsky et al.
2002). The anisotropic burning caused by the hot spot at the point of 
ignition may be revealed through the observation of burst oscillations, the
frequencies of which correspond to the NS rotation frequencies (e.g., Muno 2004). 

We searched for burst oscillations by computing 
and averaging the power spectra for the 3 bursts observed with 125 $\mu$s resolution 
event mode PCA data using a sampling rate of 2048 Hz (Nyquist frequency: 1024 Hz). 
To improve statistics, we averaged power spectra for sequential 0.25, 0.5 and
1 s sections of the light curve (4, 2, and 1 Hz resolution), with total segment lengths 
varying from 3 to 30 s. We searched in the $\sim$ 10 s rise, around the burst peak, and 
at different times during the burst decay until 65 s from the burst peak. PCA deadtime
effects cause a $\sim$ 1\% decrease in the mean value of the Poisson noise to 1.98
(from 2.0) in the Leahy normalization (see Leahy et al. 1983). We initially included 
the entire 2--60 keV energy band and made no significant detections at any point during 
the burst. We then tried varying the energy ranges included in the PCA light curves. 
With the 10--30 keV data, we made a detection 15--30 s from the burst peak around 612 
Hz (Figure 2). The significance of the signal was maximized with 0.25 s segment light 
curves; therefore, 180 separate power spectra were averaged. The signal could be 
observed in each burst separately though with less significance. We then tried expanding the 
energy range that was included in the PCA light curves; the signal weakened with the 
inclusion of either higher and lower energy photons. The 10--1024 Hz Leahy-normalized power 
spectrum is presented in Figure 2, along with a power spectrum centered on the 
signal and a histogram of the difference in power from the mean. The power at 612 Hz 
deviates from the Poisson level by 4.7$\sigma$, which with 256 bins means that there
is a 0.033\% chance that the detection is spurious. Using a power spectrum 
with 1 Hz resolution, we fit a Lorentzian function to the peak, giving a frequency $\nu_{\rm o} 
\approx$ 611.2 Hz with $\Delta \nu_{\rm FWHM} \approx 3.1$ Hz. The average rms amplitude 
of the peak is 4.8\%, which is consistent with the 5\% rms amplitude of typical burst 
oscillations observed in LMXBs (Muno 2004).
\begin{figure*} \label{power}
\centering
\includegraphics[height=5in,angle=90,viewport=0 50 502 680,clip]{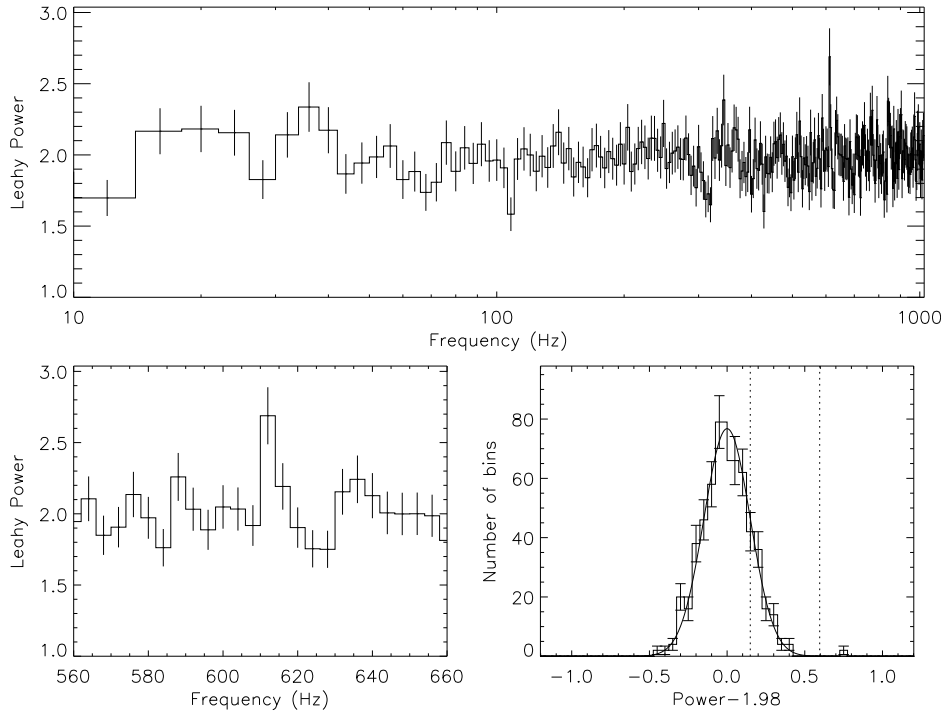}
\caption{Averaged Leahy-normalized power spectrum for the 3 bursts (15--30 s from burst
peak) observed with PCA (top and bottom left), and a histogram of the distribution of
the powers scaled to zero mean (bottom right). The frequency bins are 4 Hz. The vertical 
{\em dashed} lines in the bottom right sub-figure represent 1$\sigma$ and 4$\sigma$ standard
deviations from the mean. The power at 611.2 Hz deviates from the Poisson level by 
4.7$\sigma$.
}
\end{figure*}

\section{Persistent Emission Spectrum} \label{spec}
\subsection{Evolution Between Bursts}
Along with burst energetics and timing, we can learn much from GS 1826--238 
through its energy spectra. We began our investigation by creating spectra 
for 1 ks intervals following the onset of a burst in order to see if there were
significant changes in the best-fit parameters between bursts. Since the average 
burst recurrence time for the five complete burst intervals in the 68 ks 
\chan~observation was measured to be 12750 $\pm$ 102 s, we were able to create 1 ks
intervals up to 11--12 ks after the burst. To improve statistics we stacked 
the datasets from all five \chan~burst intervals. We also constructed nearly two 
complete burst intervals using the simultaneously acquired \rxte~data.
Although these data have time gaps, we determined the time since the previous 
burst for any individual \rxte~dataset by comparing it to the \chan~burst times. 
With the benefit of the simultaneous observations we obtained statistically 
significant measurements from 0.5--200 keV for each 1 ks interval (except 9--10 ks). 
To each dataset, we fit a model composed of Comptonized emission plus a broken 
power law. This preliminary model was only used to examine whether or not 
there were significant changes to the spectrum with time, and ultimately, a different
model was determined to be more physically plausible (see below). Nevertheless, this 
exercise showed us that the best-fit parameters remain approximately constant 
after 1 ks, although with slightly elevated soft X-ray flux through 3 ks from the 
peak of the burst. In the following fits, we therefore define the persistent spectrum to be 
for $t > 3$ ks. In the integrated persistent emission analysis below, we found 
that the inclusion of data from 1--3 ks only minimally distorted most spectral 
parameters, however the inferred $N_{\rm H}$ was underestimated by $\sim$ 50\% 
due to the excess of soft photons.

We are fortunate to have \chan~data along with \rxte~since observations of bursters 
with low-Earth orbit telescopes may miss some bursts if they occur during a gap in coverage 
when the source is occulted by the Earth or the telescope is passing through the 
South Atlantic Anomaly. Fig. 1 ({\em b}) provides an illustration of the potential 
ambiguity involved with the selection of persistent emission datasets. For example, 
it is clear that any persistent emission analysis would be slightly distorted if 
the first 3 ks of the second PCA observation (corresponding to the decay of 
the first burst observed with \chan) were included.

\subsection{Integrated Spectrum} \label{intspec}
By summing all PHA datasets that fall within the 3--12 ks interval, we further 
constrain the statistics and explore standard models for LMXBs in addition to
other models. LMXBs are typically modeled with a blackbody or disk blackbody to 
represent the NS surface or the inner accretion disk, plus Comptonized emission
which can be modeled in {\sc xspec} with {\tt comptt} or a cut-off power law 
({\tt cutoffpl}). Although these models have similar profiles, they are not 
necessarily interchangeable. While Comptonized emission can be empirically described 
with a cut-off power law, the model of Titarchuk (1994) is derived from analytical 
equations that are founded on the real analytical theory of Comptonization of soft 
photons, and the output parameters are physically explicit which allow direct 
interpretation. On the contrary, the index of a cut-off power law cannot be interpreted 
in physical terms, while in {\tt comptt} it is related to the combination of the optical 
depth and plasma temperature. Albeit, the cut-off energy of {\tt cutoffpl} 
is related to the plasma temperature as $E_{\rm cut} \sim 2 kT_{\rm e}$. For these 
reasons we chose the Titarchuk (1994) model to represent Comptonized emission. With
this description, the spectrum is governed entirely by the 
plasma temperature and the $\beta$ parameter, which characterizes the distribution law 
of the number of scatterings (i.e., $P(n) \propto \exp{(-\beta n)}$ is the probability 
that a seed photon undergoes $n$ scatterings before escaping the plasma). The optical 
depth of the plasma is calculated from the $\beta$ parameter and depends on
the input geometry: disk or spherical. Generally, for a given $\beta$ parameter, a 
larger optical depth is inferred from a spherical geometry than a disk geometry. This
is because the final spectral shape is determined by photons which undergo many more 
scatterings than the mean, and the longest dimension of a spherical plasma cloud is 
clearly shorter than that of a disk (Titarchuk 1994).
 
We initially tried to fit the data by employing a single {\tt comptt} component, but 
the resulting fit was unacceptable due to excesses above 70 keV and below 1.5 keV. The 
addition of a blackbody component reduced the soft excess, and allowed the plasma 
temperature and thus the up-scattering efficiency to increase (since $\langle 
E_{\rm f} \rangle \sim \langle E_{\rm i} \rangle e^{y}$, where $y \equiv 4kT_{\rm e} 
\tau^{2}/m_{\rm e}c^{2}$ is the Comptonization parameter) to try to match the hard 
excess. Although the fit was better ($\chi^{2}_{\nu} = 1.30$), it was still unacceptable. 
This is contrary to the results of in 't Zand et al. (1999a), who successfully used this 
model, although the 3.8 keV best-fit blackbody temperature is probably non-physical. 
The best-fit blackbody temperature that we obtained with this model was $\sim$ 0.5 keV,
which seems more reasonable.

After trying various models in conjunction with {\tt comptt}, we found the fit to be most 
improved by an additional {\tt comptt} component, resulting in a model that can be
interpreted as the Comptonization of soft photons from two distinct emission regions. 
One of the emission regions contributes 62\% of the total luminosity, so we label it as the ``primary"
emission region, and the other as the ``secondary" emission region. The primary emission region is 
characterized by $kT_{\rm s} \sim 0.3$ keV, and $kT_{\rm e} \sim$ 19.7--22.1 keV (depending
on the chosen geometry), while the secondary emission region has $kT_{\rm s} \sim 0.8$ keV and 
$kT_{\rm e} \sim$ 6.1--6.4 keV. 
The primary emission accounts for about 70\% of the flux from 0.5--3 keV, and 80\% of the 
flux from 20--200 keV, while each Comptonization component contributes nearly an equal amount 
of flux for 3--20 keV. Based on the inferred $\beta$ parameter, which is not a direct input 
in {\sc xspec}, the primary emission region has $\tau \sim 2.7$ if the region is disk-shaped, 
or $\tau \sim 6.3$ if it is spherical. The secondary emission region is characterized by 
$\tau \sim 4.7$ and $\tau \sim 9.9$ for disk and spherical geometries, respectively. 

The physical interpretation of each Comptonizing region must be inferred from the best-fit 
parameters and the corresponding flux contributions. For one, we expect the Comptonizing region 
closer to the stellar surface to have hotter seed photons, which suggests that the secondary 
component is emitted nearer to the NS. The secondary component may therefore result from 
boundary layer or surface emission. Conversely, the primary emission possibly stems from  
the accretion disk or ADC, and a disk geometry implies $\tau \sim 2.7$. From the relative flux 
contributions, we see that this interpretation requires about 3/5 of the total flux to be 
emitted from the cooler disk region. Accretion theory predicts that half of the total 
gravitational binding energy per unit mass is released in the disk, and the other half in the 
boundary layer or on the stellar surface (Frank et al. 1985). This is not inconsistent since 
the accretion disk and the boundary layer are not physically distinct; rather, there may be a 
smooth transition from one to the other. Secondly, the optical depth of the secondary 
emission (independent of the
true geometry) is large enough to naturally explain the lack of a visible blackbody component in
the GS 1826--238 spectrum. According to Inogamov \& Sunyaev (1999), the spreading of accretion 
flow from the equator to the poles leads to two bright rings of enhanced emission that are 
symmetric about the equator. The latitude of these rings increases with accretion rate, and in 
the accretion regime of GS 1826--238, i.e. $\sim$ 10$^{17}$ g s$^{-1}$ (inferred from the total 
flux), these rings lie $\sim$ 0.5--1.5 km from the equator. Therefore, assuming the blackbody
emission comes from these rings, the secondary Comptonized emission region only has to be 
large enough to cover the inner $\pm$1.5 km above and below the equator. Following in 't Zand et al. (1999b),
an approximate effective radius for the spherical emission area of the Wien seed photons is 
given by $3 \times 10^{4} d \sqrt{\frac{F_{\rm ctt}}{1+y}}/(kT_{\rm s})^{2}$ km, where $d$ is 
in kpc, the flux is measured in erg s$^{-1}$ cm$^{-2}$, and the seed photons are measured in keV. 
This approach yielded a soft photon emitting region with radius $R_{\rm s} \sim 4 
d_{\rm 6 kpc}$ km, which is only consistent with the seed photons being generated at 
the boundary layer if the region is confined to a half-thickness of $\sim$ 2.5 km, 
depending on the inclination of the system, and assuming that any emission from 
``below" the accretion disk is not observable. Finally, since the flux of photons 
passing through the boundary layer relative to the disk is larger per unit area, 
we expect Compton cooling to maintain a lower plasma temperature for the secondary 
Comptonized emission region. This is indeed what is observed (see Table 3). 

In addition to a dual Comptonization model, we also modeled the spectrum with a single 
{\tt comptt} plus a broken power law in order to give an estimate of its shape for possible 
non-thermal interpretation. If the secondary emission is indeed non-thermal, perhaps it 
stems from an ADC generated by magnetohydrodynamic turbulence, analogous to what is observed 
in the solar corona (Crosby et al. 1998). Alternatively, the emission may be synchrotron 
radiation from a relativistic jet escaping the system. Such jets have been found to be rather 
common among LMXBs and link these systems to active galactic nuclei (AGNs) (e.g., Fender 2002). 
Clearly, these possibilities are highly speculative.

\subsection{Iron Line Detection}
After obtaining a fit with the two component models, the residuals revealed a line 
feature around 6.5 keV, which prompted the addition of a Gaussian to the 
models. The best-fit value for the line was measured to be approximately $6.45$ keV, with 
a flux which corresponds to an equivalent width (EW) of $\sim$ 37.2 eV. We interpret this 
feature as a neutral Fe K$\alpha$ line. 
An F-test showed that the probability for an improvement to the fit occurring by chance is 
$5.3 \times 10^{-5}$ for the dual Comptonization model, and $6.8 \times 10^{-4}$ for the
Comptonization plus broken power law model. However, it should be noted that the use of
an F-test to measure the significance of lines may not be valid (Protassov et al. 2002).
We present the best-fit parameters for these 
models in Table 3. We also present the results of a blackbody plus cut-off power law
model, which is described below. To account for the uncertainties in the relative 
instrumental flux calibrations, we introduced a multiplicative constant into 
the spectral models. For each model, the normalizations of the \chan~instruments 
relative to PCA were $\sim 0.95$, while for HEXTE the relative normalization was 
generally $\sim$ 0.85--0.9. Figure 3 shows the spectral fit for the dual Comptonization 
model, and the residuals for all three models.
\begin{deluxetable*}{lccc} 
\tabletypesize{\scriptsize}
\tablenum{3}
\tablecolumns{4}
\tablewidth{0pt}
\tablecaption{\sc{Persistent Emission Spectral Parameters}\label{perpar}}
\tablehead{
\colhead{Model:} & 
\colhead{1. Two Comptonized} &
\colhead{2. Comptonized} &
\colhead{3. Blackbody} \\
\colhead{ } &
\colhead{(disk + spherical)\tablenotemark{a}} &
\colhead{plus Bkn. PL\tablenotemark{b}} &
\colhead{plus CPL\tablenotemark{c}}
}
\startdata
N$_{\rm H}$ ($\times 10^{21}$ cm$^{-2}$) & 1.60$^{+0.33}_{-0.61}$ & 1.50$^{+0.06}_{-0.06}$ &
4.29$^{+0.21}_{-0.19}$ \\
kT$_{\rm d}$/kT$_{\rm s}\mid_{\rm seed}$, kT$_{\rm bb}$\tablenotemark{d} (keV) & 0.42$^{+0.05}_{-0.03}$/0.82$^{+0.05}_{-0.03}$ & 0.68$^{+0.01}_{-0.01}$/0.86$^{+0.04}_{-0.03}$ & 1.34$^{+0.04}_{-0.04}$ \\ 
R$_{\rm s}$, R$_{\rm bb}$\tablenotemark{e} (km) & 4.02$^{+0.54}_{-0.37}$ & 5.88$^{+0.43}_{-0.39}$ & 2.38$^{+0.19}_{-0.18}$ \\
kT$_{\rm d}$/kT$_{\rm s}\mid_{\rm electron}$ (keV) & 20.79$^{+0.72}_{-1.08}$/6.84$^{+0.38}_{-0.33}$ & 5.85$^{+0.28}_{-0.19}$/4.74$^{+0.32}_{-0.22}$ & \nodata \\
$\tau_{\rm d}$/$\tau_{\rm s}$  & 2.56$^{+0.04}_{-0.04}$/9.38$^{+0.04}_{-0.04}$ & 4.69$^{+0.03}_{-0.04}$/10.98$^{+0.23}_{-0.18}$ & \nodata \\
$y_{\rm d}$/$y_{\rm s}$ & 1.07$^{+0.07}_{-0.09}$/4.71$^{+0.30}_{-0.27}$ & 1.01$^{+0.06}_{-0.05}$/4.47$^{+0.49}_{-0.35}$ & \nodata \\
$\Gamma_{1}$\tablenotemark{f} & \nodata & 1.19$^{+0.01}_{-0.03}$ & 1.26$^{+0.01}_{-0.02}$  \\
$\Gamma_{2}$ & \nodata & 2.43$^{+0.05}_{-0.04}$ & \nodata \\
E$_{\rm break}$, E$_{\rm cut}$ (keV) & \nodata & 35.41$^{+0.52}_{-0.47}$ & 41.58$^{+1.49}_{-1.60}$ \\ 
E$_{\rm line}$ (keV) & 6.44$^{+0.03}_{-0.12}$ & 6.45$^{+0.02}_{-0.10}$ & 6.37$^{+0.12}_{-0.11}$ \\
EW$_{\rm line}$ (eV) & 37.2$^{+10.1}_{-8.4}$ & 28.3$^{+14.3}_{-11.9}$ & 15.2\tablenotemark{g} \\
$\chi^{2}_{\nu}$ (d.o.f.) & 0.70 (521) & 0.75 (522) & 0.84 (527) \\ 
\enddata
\tablecomments{All errors are quoted at the 90\% confidence level for a 
single parameter.}
\tablenotetext{a}{{\sc xspec}: {\tt phabs*edge(comptt + comptt + gauss)}}
\tablenotetext{b}{{\sc xspec}: {\tt phabs*edge(comptt + bknpower + gauss)}}
\tablenotetext{c}{{\sc xspec}: {\tt phabs*edge(bbody + cutoffpl + gauss)}}
\tablenotetext{d}{The parameters for spherical and disk geometries are 
listed as ``disk/spherical", or ``primary/secondary" for models 1 and 2. The two 
geometries are fit together in model 1, and separately in model 2.}
\tablenotetext{e}{The spherical Wien emission radius and the blackbody radius.
The derivation of the Wien radius is described in \S~4.2.
The blackbody radius is defined by the relation $L_{\rm bb}=
4 \pi R^{2}_{\rm bb} \sigma T^{4}_{\rm eff}$, where $\sigma$ is the Stefan-Boltzmann 
constant. The blackbody normalization in {\sc xspec} is defined to be 
$L_{39}/D^{2}_{10}$, where $L_{39}$ is the blackbody luminosity in units
of 10$^{39}$ erg s$^{-1}$ and $D_{10}$ is the distance to the source in units
of 10 kpc. The derivation $R_{\rm bb}$ uses a ratio $T_{\rm bb}/T_{\rm eff}=1.4$ 
(Ebisuzaki et al. 1984, also see \S~\ref{burspec}), and a source distance of 6 kpc.}
\tablenotetext{f}{The model 2 broken power law parameters are independent of the geometry 
of the Comptonizing region.}
\tablenotetext{g}{This equivalent width is an upper limit.}
\end{deluxetable*}

\begin{figure*} \label{spectra}
\centering
\includegraphics*[width=5in,viewport=28 0 500 500,clip]{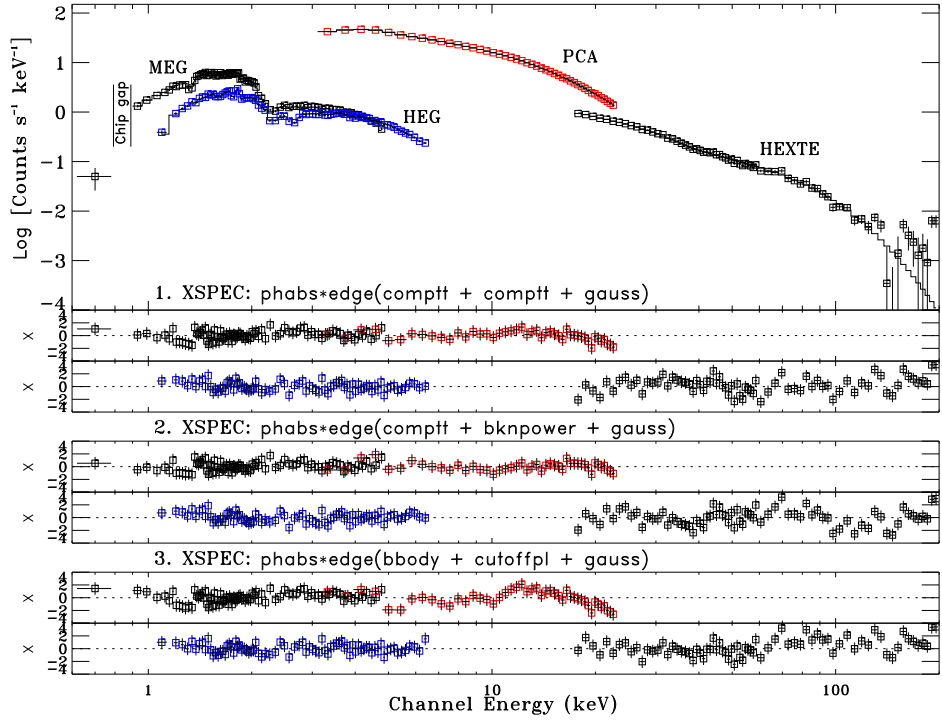}
\caption{Persistent emission spectrum and residuals of the dual Comptonization model (1), and the
residuals of models 2 and 3. For each model, the upper residuals are for the MEG and PCA, and 
the lower residuals are for the HEG and HEXTE. The range 0.8--0.9 keV is ignored since these MEG counts fall on a 
gap between ACIS CCDs. To provide clarity, we have only plotted every third (second) MEG (HEG) 
}
\end{figure*}
\subsection{Comparison to Previous Fits}
To facilitate comparison to the results of previous work, we also fitted the 
spectrum of GS 1826--238 with a blackbody plus cut-off power law model (model 3),
and a blackbody plus power law from 0.5--10 keV (not included in the table). 
These models have been used by others to provide acceptable fits to the GS 1826--238
persistent emission, but only the {\em BeppoSAX} observations had similar energy 
coverage. For example, the {\em ASCA} SIS instrument used in the Kong et al. 
(2000) analysis did not provide coverage above 10 keV, and the Barret et al. (1995)
{\em ROSAT} observation was restricted to energies below 2.2 keV. In order to compare
to the Kong et al. (2000) results, we used the same blackbody plus power law model
and omitted all data above 10 keV. With this approach we obtained a fit with 
$kT_{\rm bb} \sim 0.8$ keV and $\Gamma \sim 1.0$, which are each within $\sim$ 15\% 
of their results. Extending the energy range to 200 keV, however, revealed a higher 
blackbody temperature, a steeper photon index, and the power law had to be 
changed to a cut-off power law (model 3). This shows that the use of \rxte's 
extended energy range is crucial to minimize uncertainties in the soft component 
parameters due to less precise knowledge of the photon index or Comptonized 
component when measured with the lower \chan~energy range. 

As mentioned in \S\S~1 and 4.2, in 't Zand et al. (1999a) fit the 0.1--200 keV 
persistent emission spectrum with a blackbody plus {\tt comptt}, but they also
modeled the spectrum with a blackbody plus a cut-off power law. The best-fit blackbody 
temperature was $\sim 0.9$ keV, $\Gamma \sim 1.4$, $E_{\rm cut} \sim 52$ keV, and
$N_{\rm H} \sim 5.4 \times 10^{21}$ cm$^{-2}$, which slightly differs from $kT_{\rm bb} 
\sim 1.3$ keV, $\Gamma \sim 1.3$, $E_{\rm cut} \sim 42$ keV, and $N_{\rm H} \sim 
4.3 \times 10^{21}$ cm$^{-2}$ of model 3. If we set the model 3 parameters to those 
obtained by in 't Zand et al. (1999a), however, we obtain a similar $\chi^{2}_{\nu} 
\sim 1.4$. 

By examining the residuals, the slope of the model 3 appears to be too steep from 
$\sim$ 2.5--5 keV and 12--20 keV, and too shallow for $\sim$ 5--12 keV. A slight 
excess also remains for energies above about 60 keV (see Fig. 3 {\em bottom}). The 
2.4 km blackbody radius derived from this model is obviously smaller than expected 
for a NS, though the size of the radius may be explained if the blackbody emission 
is confined to the equatorial region. For a 10 km NS, a blackbody radius of 2.4 km 
would correspond to emission from a strip of half-height $\sim$ 0.6 km. We note
that since $E_{\rm cut} \sim 2 kT_{\rm e}$, the $\sim$ 42 keV cut-off energy is 
consistent with the $\sim$ 20 keV plasma temperature of the primary Comptonized 
component of model 1. However, unlike the dual Comptonization model and the Comptonized 
emission plus broken power law model, the addition of a line feature at $\sim$ 6.45 keV 
to model 3 did not significantly improve the fit, though we include it to preserve 
consistency. We place an upper limit on the equivalent width at 15.2 eV with 
90\% confidence.

Even though the blackbody plus cut-off power law fit that we obtained is acceptable, 
we are less confident with this model as compared to model 1 since the inferred 
$N_{\rm H}$ is inconsistent with the value obtained through the galactic hydrogen 
survey of Dickey \& Lockman (1990), where the average column density in the 
direction of GS 1826--238 was found to be $\sim$ 1.9 $\times 10^{21}$ cm$^{-2}$. 
Moreover, such a high column density is inconsistent with the X-ray halo measurements 
presented by Thompson et al. (2005, in preparation). Briefly, for the X-ray halo 
analysis we created an exposure-corrected radial profile of the source and subtracted 
a normalized model point spread function to obtain the X-ray halo profile. Using 
empirical relations from the dust scattering analysis of Draine (2003), we found 
that $\la$ 3\% of the source flux is scattered into the X-ray halo. This result was 
then used to obtain a scattering optical depth, which can be converted to an approximate
hydrogen column density through a linear regression derived using observations 
of X-ray halos around 29 sources (cf. Predehl \& Schmitt 1995, Fig. 7). The hydrogen
column density that we obtained in this fashion is consistent with the best-fit
value of models 1 and 2. By freezing $N_{\rm H}$ in model 3 to the common lower value, 
the model fits the data poorly. We therefore conclude that the common LMXB model of 
a blackbody plus Comptonized emission described by {\tt comptt} {\em or} {\tt 
cutoffpl} is not appropriate for the persistent emission spectrum of GS 1826--238.

\subsection{Interstellar Absorption Features}
The persistent emission spectrum was searched for interstellar absorption
features. This effort turned up a good candidate at about 
17.15 \AA.  The feature is broad (0.102 $\pm$ 0.027 \AA\ FWHM) with an 
optical depth at line center of 0.63 $\pm$ 0.09, when fitted with a 
Gaussian.  The formal significance is about 3.8$\sigma$; based on 
about 800 bins searched at 0.03 \AA\ binning, there is about a 6\% 
chance that one would find such a feature due to random fluctuations.  
Including a narrower feature at about 17.5 \AA\ with a significance of 
2.9$\sigma$, we modeled both features with structure in the Fe L edge due 
to the interstellar medium (ISM).  The wavelengths of these features 
are a good match to those of Fe L {\sc III} at 17.51 \AA\ and Fe 
L {\sc II} at 17.19 \AA.  The match is not perfect.  Figure 4 shows the 
near edge extended absorption fine structure (NEXAFS) of the Fe L edge 
as measured by Kortright and Kim (2000) and the edge structure of 
butadiene iron tricarbonyl (C$_7$FeH$_6$O$_3$) from the Corex data base 
maintained by A. P. Hitchcock\footnote{see 
http://unicorn.mcmaster.ca/corex}.  Neither model matches the data 
without a slight energy shift to match the L {\sc III} feature.  
Furthermore, the 17.15 \AA\ line is somewhat broader than expected and 
may have an excess of absorption at the short wavelength side.  Schulz 
et al. (2002) identified the Fe L features in the spectrum of Cyg X-1 
and suspected that there is a mix of Fe molecules, citing, in 
particular, a feature near 17.15 A that might be due to a more pure 
form of Fe.  Without the corresponding L {\sc III} feature in our 
spectrum, however, we suspect that the extra broadening might result 
from statistical fluctuation.  The models use a cosmic abundance of Fe 
($4 \times 10^{-5}$ relative to H) and the observed ISM column density 
of $2.0 \times 10^{21}$ cm$^{-2}$.  Thus, these absorption features are 
consistent with Fe in the ISM.

\begin{figure*}[!t] \label{abs}
\centering
\includegraphics[height=4.5in, angle=90]{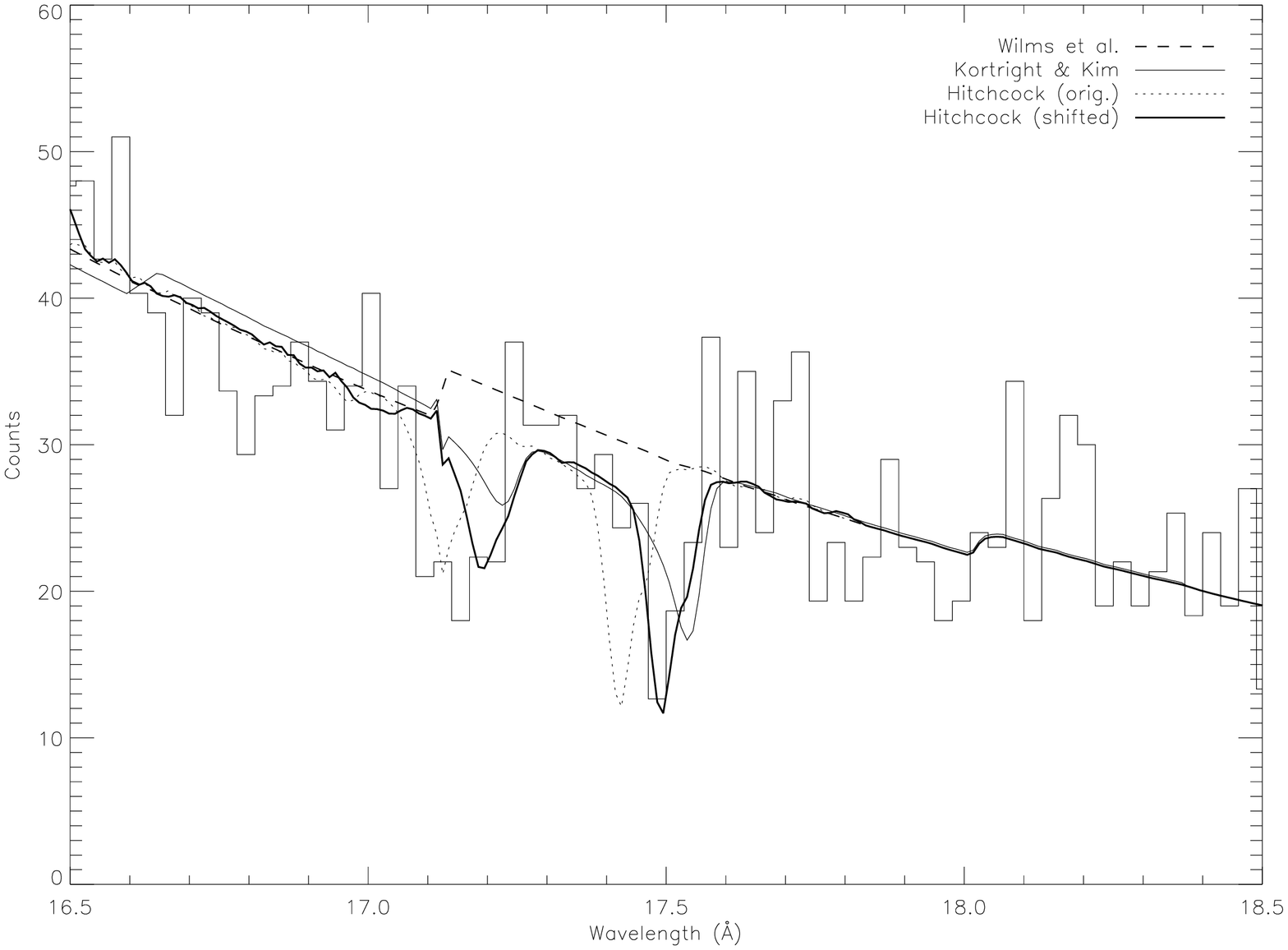}
\caption{Spectrum of the persistent emission of GS 1826--238 near 17.5 \AA.  
A global model of the continuum is shown for several models of the Fe L 
edge complex due to the ISM.  {\em Dashed curve:} The ISM opacity model 
provided in the {\tt tbabs} model of {\sc xspec}, using the 
prescription provided by Wilms et al. (2000).  {\em Thin curve:} The Fe L 
edge opacity model based on measurements by Kortright \& Kim (2000). {\em Dotted 
curve:} The Fe L edge structure is based on the transmission of 
butadiene iron tricarbonyl (C$_7$FeH$_6$O$_3$) from the Corex data base 
maintained by A. P. Hitchcock.  {\em Thick curve:} The model based on 
the Hitchcock data but shifted by 3 eV to match the data better.  The 
17.15 \AA\ line is somewhat broader than expected and may have an 
excess of absorption at the short wavelength side but these absorption 
features are generally consistent with a cosmic abundance of Fe.}
\end{figure*}

\section{Burst Spectrum and Evolution} \label{burspec}
We also studied the evolution of the spectrum throughout the burst and its decay by 
subdividing the first 1 ks from burst peak into 8 datasets with progressively larger 
integration times (Table 4). Those data above 40 keV were ignored as too few counts were obtained 
for significance. All of the datasets were fit with the {\sc xspec} model 
{\tt phabs*edge(comptt$_{\rm d}$ + comptt$_{\rm s}$ + bbody)} (d: disk, s: sphere), with the hydrogen 
column density frozen at the common value of $1.85 \times 10^{21}$ cm$^{-2}$. We initially
allowed the disk Comptonization parameters to vary, but the best-fit values for each 
dataset, including the normalizations, changed by $\la$ 5\%, so they were frozen at the 
persistent emission values. Since the blackbody emission during a burst provides a 
large influx of seed photons in the inner boundary layer, the spherical seed photon 
temperature was fixed to the blackbody temperature, which is justified since most 
Comptonized emission is emitted within $\sim$ 20 km of the surface of the NS (Frank et al. 1985). 
Table \ref{bbpar} shows the evolution of the spectral parameters.

\begin{deluxetable*}{cccccccc}
\tabletypesize{\scriptsize}
\tablenum{4}
\tablecolumns{8}
\tablewidth{0pt}
\tablecaption{\sc{Spectral Evolution During Burst Decay}\label{bbpar}}
\tablehead{
\colhead{Interval\tablenotemark{a}} &
\colhead{kT$_{\rm bb}$} &
\colhead{Norm.} &
\colhead{R$_{\rm bb}$} &
\colhead{kT$_{\rm e}$} &
\colhead{$\tau$} &
\colhead{F$_{\rm X}$: 0.5--10.0 keV} &
\colhead{0.5--40.0 keV} \\
\colhead{(s)} &
\colhead{(keV)} &
\colhead{($\times 10^{-2}$)} &
\colhead{(km)} &
\colhead{(keV)} &
\colhead{} &
\multicolumn{2}{c}{($\times 10^{-9}$ erg s$^{-1}$ cm$^{-2}$)}
}
\startdata
0--10 & 1.76$^{+0.06}_{-0.11}$ & 12.0$^{+0.6}_{-1.5}$ & 11.6$^{+0.8}_{-1.6}$ & 3.1$^{+0.6}_{-0.2}$ & 9.9$^{+22.1}_{-5.7}$ & 13.5 & 19.6 \\
10--30 & 1.72$^{+0.04}_{-0.09}$ & 10.8$^{+0.5}_{-0.5}$ & 11.5$^{+0.6}_{-1.2}$ & 3.8$^{+0.6}_{-0.5}$ & 7.8$^{+31.7}_{-4.8}$ & 10.4 & 14.9 \\
30--65 & 1.50$^{+0.09}_{-0.08}$ & 5.1$^{+0.2}_{-0.2}$ & 10.4$^{+1.3}_{-1.1}$ & 4.3$^{+0.4}_{-0.5}$ & 8.0$^{+15.6}_{-5.2}$ & 7.1 & 10.4 \\
65--100 & 1.24$^{+0.11}_{-0.11}$ & 3.1$^{+1.3}_{-1.0}$ & 11.9$^{+3.3}_{-2.8}$ & 4.5$^{+0.9}_{-0.7}$ & 10.2$^{+5.6}_{-4.1}$ & 3.8 & 6.6 \\
100--150 & 1.01$^{+0.19}_{-0.15}$ & 0.6$^{+0.7}_{-0.6}$ & 7.9$^{+5.5}_{-4.6}$ & 5.0$^{+1.4}_{-0.9}$ & 11.8$^{+6.2}_{-0.4}$ & 2.0 & 3.6 \\
150--200 & 0.86$^{+0.08}_{-0.06}$ & \nodata & \nodata & 7.1$^{+0.9}_{-2.2}$ & 11.2$^{+5.5}_{-0.7}$ & 1.4 & 2.9 \\
200--500 & 0.83$^{+0.05}_{-0.05}$ & \nodata & \nodata & 6.8$^{+1.4}_{-1.6}$ & 8.7$^{+1.1}_{-0.9}$ & 1.4 & 2.9 \\
500--1000 & 0.81$^{+0.06}_{-0.03}$ & \nodata & \nodata & 6.9$^{+0.6}_{-0.4}$ & 9.2$^{+0.6}_{-0.3}$ & 1.3 & 2.8 \\
\enddata
\tablecomments{All errors are quoted at the 90\% confidence level for
a single parameter. The {\sc xspec} model used in this analysis was {\tt phabs*edge(comptt$_{\rm d}$ 
+ comptt$_{\rm s}$ + bbody)}. Only the last two components had free parameters. The column density was 
set to $1.85 \times 10^{21}$ cm$^{-2}$, and the parameters describing disk Comptonization were fixed to the 
persistent emission best-fit values. The seed photon temperature of the spherical Comptonizing region was 
tied to the blackbody temperature. Good fits were obtained with $\chi^{2}_{\nu}$ consistently $\sim$ 0.9.}
\tablenotetext{a}{The peak of the burst is defined to take place at $t=0$.}
\end{deluxetable*}

Using the results from Table 4, we can place an upper limit to the source
distance assuming the blackbody luminosity is Eddington-limited, which it is not
since the bursts are significantly sub-Eddington. 
During the first 10 s following the peak of the burst, the average flux from 
0.5--40.0 keV was measured to be $1.96 \times 10^{-8}$ erg cm$^{-2}$ s$^{-1}$. 
We corrected the peak flux by assuming an approximately constant spectrum for this 
interval (i.e. constant spectral parameters), and by measuring the burst decay 
scale time ($\sim$ 35 s over the energy band). The \emph{peak} flux was found to be 
$\sim$ 12\% larger than the 0--10 s mean, or $2.2 \times 10^{-8}$ erg cm$^{-2}$ 
s$^{-1}$. Moreover, we must account for radiation that is absorbed or scattered
out of the line-of-sight on its way to the telescope. The hydrogen column density 
toward the source was assumed to be about $2 \times 10^{21}$ cm$^{-2}$, which 
corresponds to a scattering optical depth of 0.02 (Predehl \& Schmitt 1995). Thus 
scattering and absorption along the line of sight reduce the observed flux by 
$(1- e^{-0.02}) \approx 0.02$. 
If we assume that the blackbody normalization is also $\sim$ 12\% larger
than the 0--10 s mean, and that the \emph{unabsorbed} normalization is $\sim$ 2\% 
larger still, we arrive at a maximum source distance of 9.2 kpc\footnote{In this 
derivation we do not include the flux contributions 
from radiation that is thought to be emitted from outside the NS surface, since this
radiation would not provide an impediment to the accretion rate. Therefore, we do 
not include the Comptonized emission. (This interpretation is clearly
debatable.) The blackbody normalization is defined in Table 3 comment 
$e$. We also assume a 1.4 M$_{\odot}$ NS.} for the bursts being at the Eddington
limit. This upper limit on the distance is 
consistent with the 8 kpc upper limit derived by in 't Zand et al. (1999a).

The thermal nature of the burst spectrum with a blackbody temperature of $\sim$ 1.8 keV, 
accompanied by the cooling, is typical for a type I X-ray burst (Lewin et al. 1995). 
While the blackbody temperature deceases by a factor of about 2.2 during burst decay, the 
normalization decreases by more than a factor of 30. Nevertheless, the derived blackbody 
radius remains approximately constant as long as blackbody emission is observable. Figure 5 
shows the derived blackbody radius and the fraction of blackbody flux 
($F_{\rm bb}/F_{\rm total}$) from 2--10 keV during the first 150 s after burst ignition. 
Clearly these data are consistent with a blackbody radius of $\sim$ 10.3--11.7 km, assuming 
a source distance of 6 kpc and a ratio $T_{\rm bb}/T_{\rm eff}$ (spectral hardening factor) 
of 1.4. Such a correction must be applied since NSs do not radiate as perfect blackbodies 
during X-ray bursts because the thermalization of photons occurs at scattering optical 
depths greater than unity, where the temperature is higher than the effective 
temperature. Rather, the photons are thermalized at optical depths of $\sim$ 4--5 (e.g., 
Ebisuzaki et al. 1984, London et al. 1986, Madej et al. 2004).  Assuming that the entire 
stellar surface is involved when blackbody emission is observable, i.e. the burning does 
not cease in one area sooner than another, the slight decrease in the inferred blackbody 
radius for the 100--150 s interval (see Fig. 5) could be explained by a 
$\sim$ 10\% increase in the spectral hardening factor. This is a possibility since at 
lower effective temperatures ($T_{\rm eff} < 1.5$ keV) the spectral hardening factor 
increases with decreasing temperature, as the relative contribution of electron scattering 
to the total opacity decreases (London et al. 1986). However, within the error limits, the 
blackbody radius for this interval is still consistent with the measured values during the 
first 100 s of the burst.
\begin{figure*} \label{bbr}
\centering
\includegraphics[width=4in]{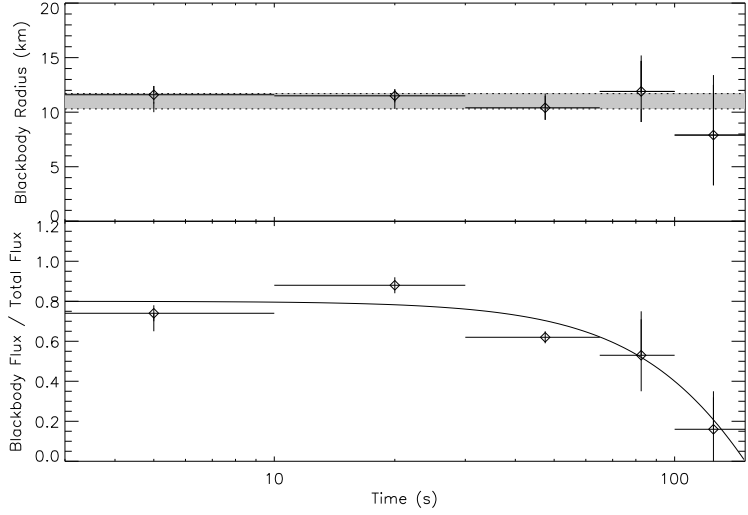}
\caption{Blackbody radius and fraction of blackbody flux ($F_{\rm bb}/F_{\rm total}$) 
from 2--10 keV during the first 150 s after burst ignition. The bottom curve is fit ``by eye'' 
with a cosine. Beyond 150 s, the {\sc xspec} blackbody normalization decays to zero. These data 
are consistent with a blackbody radius between 10.3 and 11.7 km for an assumed distance of 6 kpc.}
\end{figure*}

The derived blackbody radius cannot be interpreted as the physical size of the NS since LMXBs 
do not have isotropic radiation fields. Due to the presence of an optically thick accretion 
disk, any blackbody emission from the surface of the NS that is in the ``shadow'' of the 
accretion disk would not be directly observed but would possibly emerge as Comptonized 
emission. The only geometrical arrangement for which we can naively interpret the blackbody 
radius as the stellar radius is when the accretion disk is perpendicular to the line of sight 
($i = 0\arcdeg$). In all other cases the inferred blackbody radii underestimate the stellar 
radius, and so one must account for the covered region in order to extract a stellar radius 
from a measured blackbody temperature and flux. 

The area of the covered region increases nonlinearly with increasing binary inclination, 
moving from 0\% of the projected area when $i=0\arcdeg$ to 50\% when $i=90\arcdeg$. Furthermore, 
since accretion disks are not infinitely thin there will also be some portion of the surface 
covered by the disk. The amount of covering by the disk can be parametrized by the disk 
half-height $h$. In spite of the nonlinear relation between the amount of covering and the 
inclination and disk half-height, the fraction of the stellar surface that is covered can be 
calculated using straightforward geometrical principles. With these issues in mind we 
interpreted $\pi R_{\rm bb}^{2}$ as the observable {\em projected} area of the star, and by 
calculating the fraction of surface covered we derived an effective stellar radius 
($R_{\rm eff}$) as a function of inclination and disk half-height for source distances of 
6 kpc and 5 kpc, which is presented in Fig. 6. Using optical observations of the counterpart 
of GS 1826--238, Mescheryakov et al. (2004) were able to derive an approximate inclination 
of $i \sim$ $40\arcdeg$--$70\arcdeg$, and so we only include this range in the figure.
\begin{figure*} \label{bm}
\centering
\includegraphics[width=5in]{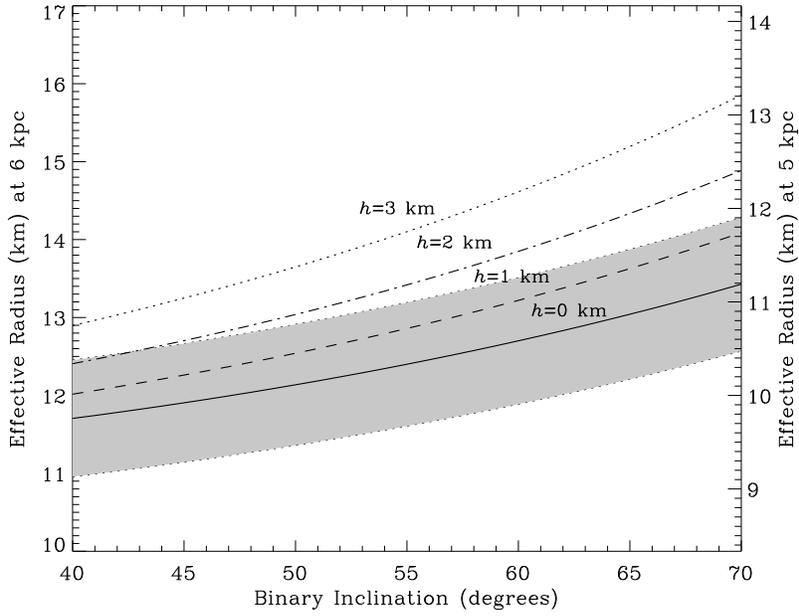}
\caption{Effective stellar radius as a function of binary inclination and 
disk half-height ($h$) for source distances of 6 kpc ({\em left axis}) and 5 kpc 
({\em right axis}). The shaded region represents the uncertainty and corresponds to 
the same range of blackbody radii as in Fig. 5. While the uncertainty is only plotted 
for $h=0$, the magnitude of the uncertainties for the other values 
of $h$ are approximately the same and have been omitted for clarity.}
\end{figure*}

From photometric measurements of the optical counterpart, Barret et al. (1995) derived an 
approximate lower limit to the distance to GS 1826--238 of 4 kpc. Assuming the range of 
possible binary inclinations used above is correct, it is apparent that a source distance 
near 5 kpc gives a NS radius within the commonly assumed range of $\sim$ 
10--12 km (see the right axis of Fig. 6). On the contrary, a distance of 6 kpc does not give 
a NS radius within the common range. Even with an infinitely thin accretion disk and an 
inclination at the lower bound ($i=40\arcdeg$) of Mescheryakov et al. (2004), the smallest 
possible stellar radius at this distance within the error limits is $\sim$ 11 km. 

From the results of the Table 4, it is clear that the burst photons immediately 
cool the Comptonizing plasma to $\sim$ 3.4 keV. Over the next 150 s, the plasma temperature 
recovers the persistent emission value of $\sim$ 6.8 keV. This type of plasma cooling during a 
burst may be an example of Compton cooling. The same inverse Compton scattering process that 
transfers energy to the persistent emission photons also transfers energy to the photons 
emitted during a burst. The difference is that during the burst the flux of photons increases 
by more than an order of magnitude (for the 0--10 s interval), so that the balance between 
heating and radiative cooling is disturbed. It is also evident that the optical depth of the 
plasma is weakly constrained in the burst spectral analysis. Although the best-fit values for 
the first few intervals seem to indicate a larger optical depth, the range of uncertainty 
still includes the persistent emission value. 
 
\section{Emission Region Geometries} \label{geo}
Since blackbody emission is observable during the bursts but not during the 
quiescent phase, we can speculate as to the geometries of the corresponding 
emission regions. To begin, GS 1826--238 has a weak magnetic field so infalling 
matter probably impacts the NS along the equator rather than at the poles. The 
release of gravitational binding energy through thermalization of the accreted 
material may result in a strip of blackbody emission along the equator (Church 
\& Ba\l uci\'{n}ska-Church 2001), or the accumulation and spreading of accretion 
flow from the equator toward the poles may cause two bright emission rings that 
are symmetric about the equator (Inogamov \& Sunyaev 1999). The inability to see 
blackbody emission in the interval between bursts suggests that the equatorial 
strip or enhanced bright rings must be covered by an optically thick layer. This 
is explained naturally by the persistent emission dual Comptonization model. 
 
At the onset of a burst, ignition likely starts near the equator and then rapidly 
spreads to cover the entire stellar surface in a couple seconds or less (Spitkovsky 
et al. 2002). Once the entire accumulated layer is burning, the subsequent blackbody 
emission indicates that there is an optical path for the radiation outside of the 
equatorial strip. It could be argued that the accretion flow is disrupted during 
a burst, or that a spherical corona surrounding the NS is temporarily blown away,
and that this provides a path for the blackbody emission. However, the flux provided 
by the inner Comptonizing region also increases substantially during a burst, so the 
boundary layer plasma must still be present to up-scatter the seed photons. Moreover, 
our measurement of the Wien emission radius ($\sim$ 4 km) is too small for the corona
to completely surround the star. Instead, we assume that this emission region is likely 
confined to the equatorial region. This explanation supports theories where the 
Comptonizing plasma or ADC is geometrically thin.

\section{Summary}
Our investigation of the LMXB GS 1826--238 using simultaneous \rxte~and \chan~observations 
has led us to the following conclusions. From the observation of five uninterrupted burst 
intervals with \chan~in July 2002, we measured a burst recurrence time of 3.54 $\pm$ 
0.03 hr, which is consistent with G04's measurement. G04's measured decrease in the burst 
recurrence time between 1997 and 2002 has been coupled with an even larger percentage increase 
in the mean persistent flux, possibly indicating that a greater fraction (as compared to 1997) 
of the stellar surface may be covered prior to ignition. We detected 611.2 Hz burst 
oscillations with a 0.033\% chance that the signal resulted from random 
fluctuations. The average rms amplitude of the peak is 4.8\%.

The \emph{RXTE/Chandra} 0.5--200 keV integrated persistent emission spectrum is best fit 
with a dual Comptonization model, whereby two distinct Comptonizing regions exist and 
are characterized by a different set of parameters. The extended energy range of \rxte~is 
essential to constraining the photon index or Comptonization parameter, and eliminates the 
need for a blackbody component. This result is contrary to most spectral models of LMXBs that
include a visible blackbody component. The spectrum also requires a neutral Fe K$\alpha$
emission line at 6.45 keV with EW $\sim$ 37.2 eV. In addition, we find strong 
evidence of interstellar Fe L absorption features at the about 17.15 \AA~and 17.5 \AA, with 
significance 3.8$\sigma$ and 2.9$\sigma$, respectively.

During a burst, blackbody emission accounts for the majority of the flux though it
quickly disappears after $\sim$ 150 s. Throughout this period, the data are consistent with a 
blackbody radius between 10.3 km and 11.7 km for a distance of 6 kpc if one assumes the
blackbody flux comes from the full $4\pi R^{2}$ of the neutron star surface. By accounting 
for the fraction of the surface that is obscured by the accretion disk, however, the source 
distance must be nearer to 5 kpc to be consistent with a neutron star radius of 
10--12 km. We also see Compton cooling during the bursts, as the plasma temperature 
immediately decreases to $\sim$ 3 keV and then slowly returns to the persistent emission
value of $\sim$ 6.8 keV after about 150 s. Since blackbody emission is not observed in the 
persistent spectrum yet dominates the burst spectrum, we conclude that the emission from 
those regions of the stellar surface along the equator are covered in the persistent phase. 
During the burst the entire surface is radiating as a blackbody, and so this emission can be 
seen outside of the covered region. 

\acknowledgements
We gratefully acknowledge support from NASA grants NAS5-30702 and GO2-3060X. 
We thank Tod Strohmayer for identifying a mistake we initially made when creating
power spectra in the search for burst oscillations, and Lev Titarchuk for useful
information regarding his {\tt comptt} model in {\sc xspec}. We also appreciate
the input given by J\"{o}rn Wilms, and many comments and suggestions by the referee
which significantly improved this paper. This research made use of data 
the High Energy Astrophysics Science Archive Research Center Online Service, provided 
by the NASA/Goddard Space Flight Center. The analysis of the \chan~data
made extensive use of the Chandra Interactive Analysis of Observations (CIAO),
\url{http://cxc.harvard.edu/ciao}.

\pagebreak

\end{document}